\documentclass[draftclsnofoot,onecolumn,11pt]{IEEEtran}

\usepackage{amsmath}
\usepackage{cases}
\usepackage{amsfonts}
\usepackage{amssymb}
\usepackage{boxedminipage}
\usepackage{graphicx}
\usepackage[usenames]{color}
\usepackage{array,color}
\usepackage{colortbl}
\usepackage{lineno}
\usepackage{slashbox,pict2e}
\usepackage{stfloats}
\usepackage{algorithm}
\usepackage{algorithmic}
\usepackage{multicol}
\usepackage{booktabs}
\usepackage{makecell}
\usepackage{bm}
\usepackage{url}
\usepackage{cite}

\newcommand\blfootnote[1]{%
  \begingroup
  \renewcommand\thefootnote{}\footnote{#1}%
  \addtocounter{footnote}{-1}%
  \endgroup
}

\begin{document}

\baselineskip=24pt

\title{ Mobile Millimeter Wave Channel Acquisition, Tracking, and Abrupt Change Detection}
\author{Chuang Zhang$^*$, ~Dongning Guo$^\dag$, ~Pingyi Fan$^*$ \\
$^*$Department of Electronic Engineering, Tsinghua University, Beijing, P.R. China\\
$^\dag$Department of Electrical Engineering and Computer Science, Northwestern University, Evanston, IL 60208, USA\\
E-mail:~zhangchuang11@mails.tsinghua.edu.cn, ~dGuo@northwestern.edu, ~fpy@tsinghua.edu.cn}
\maketitle

\begin{abstract}
Millimeter wave provides a promising approach for meeting the ever-growing traffic demand in next generation wireless networks. It is crucial to obtain relatively accurate channel state information so that beamforming/combining can be performed to compensate for severe path loss in this band. In contrast to lower frequencies, a typical mobile millimeter wave channel consists of a few dominant paths. It is generally sufficient to estimate the path gains, angles of departure (AoD), and angles of arrival (AoA) of those paths. In this paper, multiple transmit and receive antennas and beamforming with a single baseband processing chain are assumed. We propose a framework for estimating millimeter wave channels with intermittent abrupt changes (e.g., blockage or emergence of dominant paths) and slow variations of AoDs and AoAs. The solution consists of three components: tracking of the slow channel variations, detection of abrupt changes, followed by (re-)acquisition of channel (and back to the tracking stage). For acquisition, we formulate a least squares problem and find its solution based on the Levenberg-Marquardt algorithm. To track slow variations of AoDs and AoAs, we propose a new approach using Kalman filtering. Finally, an algorithm based on a likelihood test is devised for detecting abrupt changes.  Simulation results show that, with moderate signal-to-noise ratios, the proposed scheme can achieve more than 8 dB higher estimation accuracy than several other methods using the same number of pilots.
\end{abstract}


\blfootnote{This work was supported in part by a gift from Futurewei Technologies, and was presented in part at the 2016 International Conference on Communications \cite{chuang2016tadammwc}.}

\section{Introduction} \label{sec_intro}


Millimeter wave (mm-wave) communication is a promising technique for meeting the ever-increasing mobile traffic demand in next generation wireless communication systems due to vast swaths of available spectrum \cite{boccardi2014fdtd5g, andrews2014ww5gb, rappaport2013mwmc5gc, pi2011aimwmbs}. To utilize this band, it is essential to obtain the propagation characteristics and establish proper channel models. Reference \cite{rappaport2015wideband} gave an extensive summary of mm-wave propagation measurements and corresponding large scale channel models. Reference \cite{samimi201628} presented small-scale measurement results and established some small-scale spatial fading models. Based on this, \cite{samimi2016mimo} further extended the mm-wave single input single output (SISO) modeling approach to multi input multi output (MIMO) case and generated the power delay profiles for mm-wave MIMO channel. Differently from \cite{rappaport2015wideband, samimi201628, samimi2016mimo}, which focused on channel measurements and modeling in urban scenarios, \cite{degli2014ray-tracing-based} established a 3-D ray tracing model for indoor scenario based on measurements. As indicated by these measurements, attenuation loss in mm-wave band is relatively high, thus, directional beamforming and combining should be applied in order to attain sufficient signal-to-noise ratio (SNR), which generally requires accurate channel state information (CSI) at the transmitter and receiver.

Channel estimation in wideband mm-wave communication differs from that in lower frequencies in two major aspects. First, due to high cost and power consumption of analog-to-digital conversion at extremely high sampling rates, transmitters and receivers can only be equipped with a limited number of radio frequency (RF) chains \cite{ayach2014sspmwms}. This imposes some constraints on the type of beamforming/combining that can be employed. Second, mm-wave channel has limited scatterings due to directionality, large attenuation loss, and high absorption loss \cite{akdeniz2014mmwcmcce}. Therefore, it is sufficient to estimate channel parameters of these limited scattering paths instead of each element of a large channel matrix.

One straightforward approach to estimate the mm-wave channel is by searching the angle of departures (AoDs) and angle of arrivals (AoAs) of the scattering paths exhaustively, and use the direction with the largest gain as the beamforming/combining direction, as proposed in the IEEE 802.11ad standard. However, the estimation accuracy of this approach is not high since it only obtains one direction of the channel. There are other schemes which can achieve higher estimation accuracy, like the ones in \cite{alkhateeb2014cehpmwcs} \cite{lee2014essechmsmwc}. In \cite{alkhateeb2014cehpmwcs}, the authors proposed an adaptive algorithm to estimate the channel, which essentially searches paths using beamforming vectors with different beamwidths in different stages. In \cite{lee2014essechmsmwc}, the authors formulated the channel estimation problem as a sparse signal recovery problem and solved it with the orthogonal matching pursuit (OMP) algorithm. Those algorithms either rely on transceivers with several RF chains \cite{alkhateeb2014cehpmwcs} or a large number of quantization levels for AoD and AoA \cite{lee2014essechmsmwc}. For more practical transceivers with a single RF chain and limited quantization levels of phase shifters, their improvements over the exhaustive search approach are limited.

We also note that variations of a mobile mm-wave channel typically takes two different forms: The gain and phase of a given path vary in a continuous fashion, which is generally trackable. A new path may appear at any time and then disappear at a later time due to blockage. If abrupt channel changes (like blockage or emergence of dominant paths) do not occur often, then it is only necessary to detect abrupt changes and obtain an initial estimate after such a detection until the next abrupt change, a tracking approach using less frequent pilots can be sufficient.

This work studies the mm-wave channel estimation problem. The main contributions are:

\begin{figure}
  \centering
  \includegraphics[width=0.4\columnwidth]{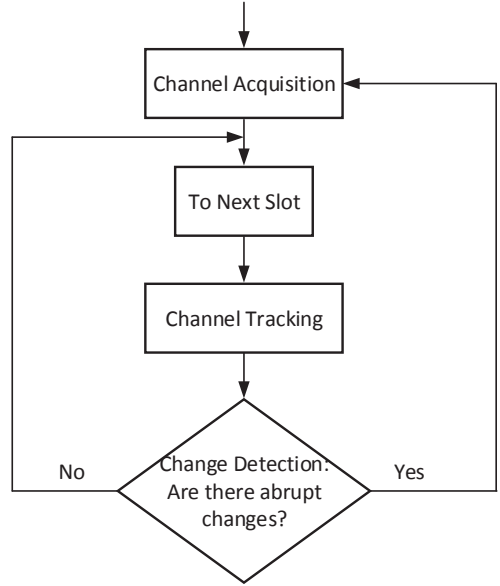}\\
  \caption{Channel acquisition, tracking, and abrupt change detection system.}\label{fig_gen_procedure}
\end{figure}

1. We propose a dual timescale mm-wave channel variation model to characterize slow variations of AoDs and AoAs and intermittent abrupt changes. We treat channel estimation as three integrated components: channel acquisition, tracking, and abrupt change detection as depicted in Fig. \ref{fig_gen_procedure}.

2. We propose an efficient channel acquisition algorithm for obtaining initial CSI. The algorithm is based on least squares and requires fewer pilots than the existing schemes in \cite{alkhateeb2014cehpmwcs} \cite{lee2014essechmsmwc} to achieve the same estimation performance.

3. We devise an algorithm for tracking slow variations of AoDs and AoAs using a Kalman filter. This method can maintain high tracking accuracy with relatively low overhead even under low-SNR conditions.

4. We develop an abrupt change detection method based on the Kalman filter tracking method.

5. We integrate the proposed channel acquisition, tracking, and abrupt change detection schemes. Simulations are carried out to evaluate each proposed algorithm independently and together. The results demonstrate the superior performance of the proposed scheme over existing ones.

The rest of the paper is organized as follows. In Section \ref{sec_sysmod}, we present the system model. In Section \ref{sec_transpolicy}, the transmission policy regarding how to send pilots is discussed. In Section \ref{sec_chacq}, we propose the channel acquisition method based on least squares. In Section \ref{sec_chtrack}, we introduce the Kalman filter based tracking method. In Section \ref{sec_acd}, the abrupt change detection approach is presented. In Section \ref{sec_sysscheme}, we discuss the integrated schemes and design of pilots. Simulation results are presented in Section \ref{sec_simulation} and conclusion is drawn in Section \ref{sec_conclusion}.

Throughout this paper, the following notations will be used. Matrices are denoted by bold uppercase letters (e.g., $\mathbf{A}$), vectors are denoted by bold lowercase letters (e.g., $\mathbf{a}$), scalars are denoted by lowercase letters (e.g., $a$). The transpose, conjugate, Hermitian (conjugate transpose), and pseudo-inverse of matrix $\mathbf{A}$ are denoted as $\mathbf{A}^T$, $\mathbf{A}^*$, $\mathbf{A}^H$, and $\mathbf{A}^\dag$, respectively. An $N\times N$ identity matrix is denoted as $\mathbf{I}_{N}$.  The Kronecker product of $\mathbf{A}$ and $\mathbf{B}$ is denoted as $\mathbf{A}\otimes\mathbf{B}$. The matrix formed by getting the real (resp., imaginary) part of each element in $\mathbf{A}$ is denoted as $\text{Re}(\mathbf{A})$ (resp., $\text{Im}(\mathbf{A})$). The absolute value is denoted as $|*|$. The $p$ norm is denoted as $||*||_p$.

\section{System Model} \label{sec_sysmod}

\subsection{Structure of the transmitter and receiver} \label{subsec_strtransreceiv}

\begin{figure}
  \centering
  \includegraphics[width=0.7\columnwidth]{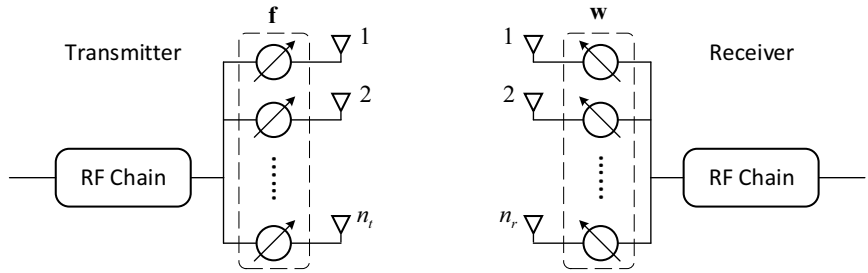}\\
  \caption{Transmitter and receiver.}\label{fig_txrx1RF}
\end{figure}

We assume that both the transmitter and the receiver utilize uniform linear arrays (ULAs) with half-wavelength antenna spacing. (The model and results easily generalize to arbitrary antenna separations.) As shown in Fig. \ref{fig_txrx1RF}, the transmitter has $n_t$ antennas and the receiver has $n_r$ antennas. For ease of implementation, we assume that both the transmitter and the receiver have a single RF chain, so that only analog beamforming/combining can be applied. Moreover, we assume that each element in the beamforming/combining vectors is of variable phase and constant amplitude. Besides, each phase shifter can only have a small number of quantization levels at the transmitter and the receiver, respectively.


\subsection{Channel model} \label{subsec_chamod}

\begin{figure}
  \centering
  \includegraphics[width=0.7\columnwidth]{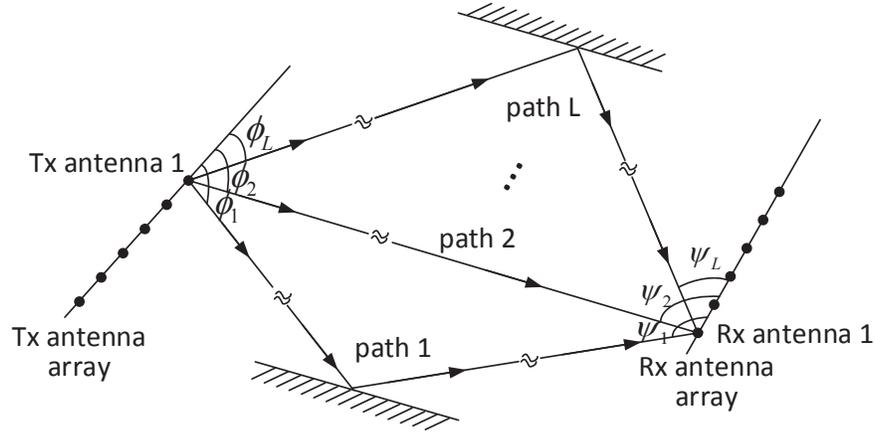}\\
  \caption{L-scatterer channel model.}\label{fig_chmod}
\end{figure}

As illustrated in Fig. \ref{fig_chmod}, we adopt the channel model from \cite{alkhateeb2014cehpmwcs} \cite{lee2014essechmsmwc}. Suppose there are up to $L$ paths, let $\phi_{l}, \psi_{l}$ be the AoD, AoA of path $l$, $l=1,\ldots,L$. Also define
\begin{flalign} \label{eqn_AoDvec}
\mathbf{e}_t(\phi)&=\tfrac{1}{\sqrt{n_t}}[1,e^{-j\pi\cos{\phi}},\ldots,e^{-j\pi(n_t-1)\cos{\phi}}]^T,\\
\mathbf{e}_r(\psi)&=\tfrac{1}{\sqrt{n_r}}[1,e^{-j\pi\cos{\psi}},\ldots,e^{-j\pi(n_r-1)\cos{\psi}}]^T.\label{eqn_AoAvec}
\end{flalign}
Then the $L$-scatterer channel can be expressed as  \cite[P.311]{tse2005fowc}
\begin{align}\label{eqn_chLscatterers}
\mathbf{H}=\sum_{l=1}^{L}\alpha_l\mathbf{e}_r(\psi_{l})\mathbf{e}_t^H(\phi_{l}),
\end{align}
where
\begin{align}\label{eqn_pathgainl}
\alpha_l=\rho_l\sqrt{n_tn_r}e^{-j\frac{2\pi d_{l}}{\lambda_c}}
\end{align}
is the gain of path $l$, $\rho_l$ and $d_l$ are the attenuation and distance between transmit antenna $1$ and receive antenna $1$ along path $l$, respectively, and $\lambda_c$ is the carrier wavelength.
Let
\begin{align}
\bm{\alpha}&=[\alpha_1,\ldots,\alpha_L]^T, \label{eqn_alpha}\\
\bm{\theta}&=[\phi_{1},\ldots,\phi_{L},\psi_{1},\ldots,\psi_{L}]^T. \label{eqn_theta}
\end{align}
Then the path gain vector $\bm{\alpha}$ and the angle vector $\bm{\theta}$ fully determine the channel.

\subsection{Channel variation model} \label{subsec_chavarmod}

We introduce a simple mm-wave channel model to capture two types of variations: 1) abrupt changes due to sudden environmental change, like the blockage of an existing path or appearance of a new path with significant gain; 2) continuous changes in the AoDs and AoAs of current paths, e.g., due to rotation of the mobile device or vibrations of base station poles \cite{hur2013mmbwbascn}.

Let time be slotted, where each slot consists of a fixed number of symbols. First, let the arrivals of paths $1,2,\dots$ form a homogeneous Bernoulli process at the slot level, where the durations of those paths in slots are independent  geometric random variables. In fact, the paths can be viewed as customers to an M/M/$\infty$ queue. Arrival of a new path and departure of an existing one are both referred to as an \emph{abrupt change}. We refer to the period between any two consecutive abrupt changes as a \emph{block}, and each block consists of a random number of slots.

Second, the AoA and AoD of a new path are assumed to be uniformly distributed upon arrival and then vary slowly until the path departs, while its gain remains constant. Specifically, the angle vector defined in (\ref{eqn_theta}) varies according to
\begin{align} \label{eqn_anglevary}
\bm{\theta}(n)=\bm{\theta}(n-1)+\mathbf{u}(n),
\end{align}
where $n$ denotes the $n$th slot, $\mathbf{u}(n)\sim \mathcal{N}(0,\mathbf{Q}_u)$ is Gaussion noise, and $\mathbf{Q}_u$ is a $2L\times 2L$ covariance matrix.

On the block timescale, we determine when abrupt changes occur and subsequently acquire the parameters of dominant paths. On the slot timescale, with the initial estimate of the channel, we track the slow variations of those paths.

\section{Beamforming and combining vectors} \label{sec_transpolicy}

\begin{figure}
  \centering
  \includegraphics[width=0.8\columnwidth]{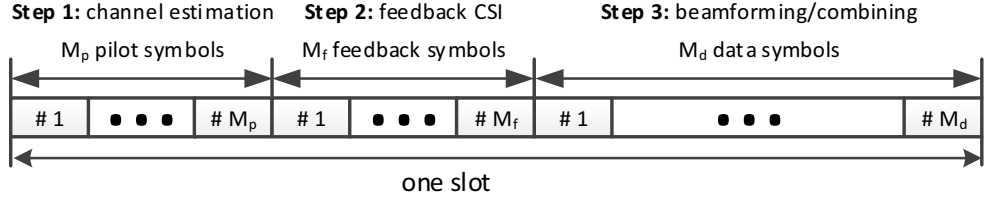}\\
  \caption{Structure of a slot.}\label{fig_frame}
\end{figure}

As shown in Fig. \ref{fig_frame}, each slot is divided into three parts. In the first part, the transmitter sends pilots and the receiver estimates the channel. In the second part, the receiver feeds back the CSI, and in the third part, the transmitter and the receiver use appropriate beamforming and combining vectors respectively to transmit data. If the transmitter sends the beamforming vector $\mathbf{f}$, and the receiver combines with vector $\mathbf{w}$, then the observation at the receiver is
\begin{align} \label{eqn_observation}
y=&\mathbf{w}^H\mathbf{H}\mathbf{f}+\mathbf{w}^H\mathbf{z},
\end{align}
where $\mathbf{z}\sim \mathcal{CN}(0,\sigma_z^2\mathbf{I}_{n_r})$ is Gaussian noise.

Suppose we use $\mathbf{f}=\mathbf{e}_t(\bar{\phi})$ and $\mathbf{w}=\mathbf{e}_r(\bar{\psi})$, where the channel is described by (\ref{eqn_chLscatterers}). Then it can be calculated that the beamforming and combining gains of path $l$ are
\begin{flalign}
\mathbf{e}_t^H(\phi_{l})\mathbf{e}_t(\bar{\phi})&=\frac{1}{n_t}\sum_{i=0}^{n_t-1}e^{j\pi(\cos{\phi_l}-\cos{\bar{\phi}})}                                            \\
&=\frac{1}{n_t}\frac{1-e^{j\pi n_t(\cos{\phi_l}-\cos{\bar{\phi}})}}{1-e^{j\pi(\cos{\phi_l}-\cos{\bar{\phi}})}} \label{eqn_fsample}
\end{flalign}
and
\begin{flalign}\label{eqn_wsample}
\mathbf{e}_r^H(\bar{\psi})\mathbf{e}_r(\psi_{l})
=\frac{1}{n_r}\frac{1-e^{-j\pi n_r(\cos{\psi_l}-\cos{\bar{\psi}})}}{1-e^{-j\pi(\cos{\psi_l}-\cos{\bar{\psi}})}},
\end{flalign}
respectively \cite[Chapter 7]{tse2005fowc}. Since the cosine function is symmetric around $\pi$, we only need to consider the range $[0, \pi]$ when estimating the angles. For instance, if $\phi_l=\frac{4\pi}{3}$ and the estimate of $\phi_l$ is  $\frac{\pi}{3}$, then based on (\ref{eqn_fsample}), the estimate can be regarded as accurate.

Most work (e.g., \cite{alkhateeb2014cehpmwcs} \cite{lee2014essechmsmwc}) quantizes the range $[0,\pi]$ uniformly. However, it leads to better results to quantize the range of $\cos\phi_l$, i.e., $[-1,1]$ uniformly and then use the angle corresponding to the center of each bin  as the beamforming direction. We will discuss the quantization issue and the required number of pilots further in Section \ref{sec_sysscheme}.

Let $M_p=m_rm_t$ pilots be transmitted in total, we select $m_t$ pilot beamforming directions $\bar{\phi}_{p}$, $p=1,\ldots,m_t$ in $[0,\pi]$, and $m_r$ pilot combining directions $\bar{\psi}_q$, $q=1,\ldots,m_r$ in $[0,\pi]$. Each pilot beamforming vector $\mathbf{f}_p=\mathbf{e}_t(\bar{\phi}_p)$, $p=1,\dots,m_t$, is repeated $m_r$ times, so that it can be received using all $m_r$ pilot combining vectors, $\mathbf{w}_q=\mathbf{e}_r(\bar{\psi}_q)$, $q=1,\ldots,m_r$. The corresponding received symbols are expressed as
\begin{align} \label{eqn_obValue}
y_{qp}=\mathbf{w}_q^H\mathbf{H}\mathbf{f}_p+\mathbf{w}_q^H\mathbf{z}_{qp},
\end{align}
where $\mathbf{z}_{qp}\sim\mathcal{N}(0,\sigma_z^2\mathbf{I}_{n_r})$.

Let $\mathbf{Y}$ denote the $m_r\times m_t$ matrix $(y_{qp})$,  $\mathbf{W}=[\mathbf{w}_1,\mathbf{w}_2,\ldots,\mathbf{w}_{m_r}]$, and $\mathbf{F}=[\mathbf{f}_1,\mathbf{f}_2,\ldots,\mathbf{f}_{m_t}]$, then the observed data can be written as
\begin{align} \label{eqn_obserMatrix}
\mathbf{Y}=\mathbf{W}^H\mathbf{H}\mathbf{F}+\mathbf{V},
\end{align}
where $\mathbf{V}$ is a $m_r\times m_t$ noise matrix with independent identically distributed (i.i.d.) Gaussian entries.

We vectorize the observation matrix by concatenating the columns of $\mathbf{Y}$. Let $\mathbf{y}=\text{vec}(\mathbf{Y})$, $\mathbf{v}=\text{vec}(\mathbf{V})$ and define $G_l^{qp}(\bm{\theta})=\mathbf{e}_r^H(\bar{\psi}_{q})\mathbf{e}_r(\psi_{l})\mathbf{e}_t^H(\phi_{l})\mathbf{e}_t(\bar{\phi}_{p})$ as the beamforming and combining gain of path $l$ with $\mathbf{e}_t(\bar{\phi}_{p})$ and $\mathbf{e}_r(\bar{\psi}_{q})$. Let $\bm{\Phi}(\bm{\theta})$ be an $m_rm_t\times L$ matrix with $G_l^{qp}(\bm{\theta})$ as its entries,
\begin{align}
\bm{\Phi}_{q+(p-1)m_r,l}(\bm{\theta})=G_l^{qp}(\bm{\theta}),
\end{align}
then
\begin{align} \label{eqn_vecobserve}
\mathbf{y}=\bm{\Phi}(\bm{\theta})\bm{\alpha}+\mathbf{v},
\end{align}
where $\mathbf{v}\sim \mathcal{C}N(0, \sigma^2_v\mathbf{I}_{m_rm_t})$, and $\sigma^2_v=\sigma^2_z$.

\section{Channel acquisition} \label{sec_chacq}

In this section, we propose an algorithm to estimate the channel parameters, i.e., the path gain vector $\bm{\alpha}$ and the angle vector $\bm{\theta}$. The estimation problem can be formulated as
\begin{align}\label{eqn_lsori}
\min_{\bm{\alpha},\bm{\theta}}\quad ||\mathbf{y}-\bm{\Phi}(\bm{\theta})\bm{\alpha}||_2^2.
\end{align}
We can obtain an explicit expression of $\bm{\alpha}$ for a given $\bm{\theta}$ with $\bm{\alpha}=\bm{\Phi}^\dag(\bm{\theta})\mathbf{y}$ \cite[p. 256]{kay2009estimation}. Then problem (\ref{eqn_lsori}) can be reformulated as
\begin{align}\label{eqn_lsrealvartheta}
\min_{\bm{\theta}}\quad ||\mathbf{y}-\bm{\Phi}(\bm{\theta})\bm{\Phi}^\dag(\bm{\theta})\mathbf{y}||_2^2.
\end{align}

\subsection{The Levenberg-Marquardt algorithm}

We propose to use the Levenberg-Marquardt (LM) algorithm  \cite{levenberg1944amscnpls}-\cite{mor1978LMait} to solve the nonlinear least squares problem (\ref{eqn_lsrealvartheta}). Let
\begin{align} \label{eqn_rfunc}
\mathbf{r}(\bm{\theta})=\big(\mathbf{I}-\bm{\Phi}(\bm{\theta})\bm{\Phi}^\dag(\bm{\theta})\big)\mathbf{y}.
\end{align}
The LM algorithm is iterative with the following steps.

Step 0. Pick some $\bm{\theta}_0$ as the starting point and let $k=1$;

Step 1. At iteration $k$, linearize $\mathbf{r}(\bm{\theta})$ w.r.t. $\bm{\theta}$ in an ellipsoid around $\bm{\theta}_{k-1}$, i.e., let
\begin{align} \label{eqn_lswc}
\bar{\mathbf{r}}(\bm{\theta})=\mathbf{r}(\bm{\theta}_{k-1})+ \mathbf{J}(\bm{\theta}_{k-1})(\bm{\theta}-\bm{\theta}_{k-1}), \quad ||\mathbf{S}\bm{\theta}||_2\leq \delta,
\end{align}
where $\mathbf{J}(\bm{\theta}_{k-1})=\frac{\partial \mathbf{r}}{\partial \bm{\theta}}|_{\bm{\theta}=\bm{\theta}_{k-1}}$ is the Jacobian matrix of $\mathbf{r}(\bm{\theta})$ at $\bm{\theta}_{k-1}$, $\mathbf{S}$ is some nonsingular matrix, which is diagonal, and $\delta$ is selected to make the linear approximation accurate;

Step 2. Based on the linearized $\bar{\mathbf{r}}(\bm{\theta})$, solve the following linear least squares problem
\begin{align}
\begin{array}{ll}
\min\limits_{\bm{\theta}} & \quad ||\bar{\mathbf{r}}(\bm{\theta}_{k-1})+ \mathbf{J}(\bm{\theta}_{k-1})(\bm{\theta}-\bm{\theta}_{k-1})||_2^2  \\
\text{s.t.} & \quad ||\mathbf{S}\bm{\theta}||_2\leq \delta,
\end{array}
\end{align}
and call the optimal solution $\bm{\theta}_k$;

Step 3. Repeat steps 1, 2 until the difference between $\bm{\theta}_{k}$ and $\bm{\theta}_{k-1}$ is smaller than a predefined threshold or the approximation error in the linearization of $\mathbf{r}(\bm{\theta})$ is smaller than a predefined threshold.

The following two issues have to be resolved in order to apply the LM algorithm to problem (\ref{eqn_lsrealvartheta}).

1. Because $\mathbf{r}(\bm{\theta})$ is not convex w.r.t. $\bm{\theta}$, there can be many local optima. Therefore it is necessary to obtain a starting point which would converge to a good local optimum.

2. The calculation of the Jacobian matrix $\mathbf{J}$ is nontrivial since $\mathbf{r}(\bm{\theta})$ involves pseudo-inverse.

We devote the following two subsections to solve the above two issues.

\subsection{To obtain a good starting point} \label{subsec_toagsp}

We reformulate the vectorized observation $\mathbf{y}$ as follows
\begin{align} \label{eqn_vecobs2multiacc}
\mathbf{y}&=\text{vec}(\mathbf{Y})  \\
&= \big(\mathbf{F}^T\otimes \mathbf{W}^H\big)\text{vec}(\mathbf{H})+\mathbf{v} \label{eqn_vecobs2multiacc_a}\\
&= \sum_{l=1}^L\Big(\big(\mathbf{F}^T\otimes \mathbf{W}^H\big) \big(\mathbf{e}^*_t(\cos{\phi_l})\otimes\mathbf{e}_r(\cos{\psi_l})\big)\Big)\alpha_l +\mathbf{v}, \label{eqn_vecobs2multiacc_b}
\end{align}
where both $(\ref{eqn_vecobs2multiacc_a})$ and $(\ref{eqn_vecobs2multiacc_b})$ follow from the identity $\text{vec}(\mathbf{A}\mathbf{B}\mathbf{C})=\big(\mathbf{C}^T\otimes\mathbf{A}\big)\text{vec}(\mathbf{B})$ \cite[Theorem 13.26]{laub2004mafse}.

The problem is to estimate $\phi_l$, $\psi_l$, and $\alpha_l$ for each path $l$. We use successive interference cancellation. We first find the path with the largest gain, and use the corresponding $\bar{\phi}_p$ and $\bar{\psi}_q$ as approximations for $\phi_1$ and $\psi_1$, respectively. Since we do not know the channels of other paths,  we simply use matched filter for detecting the path gain $\alpha_1$. Then we subtract the influence of path 1 based on the estimate of $\alpha_1, \phi_1, \psi_1$ and repeat this process to detect the remaining paths. A starting point of $\bm{\theta}_0$ is thus obtained in the form of (\ref{eqn_theta}). Since the number of paths $L$ in the channel is unknown in practice, we set a threshold $\alpha_{\text{th}}$ for the path gain and only keep paths with gains larger than the threshold. The preceding procedure is summarized as Algorithm \ref{alg_sivartheta}.

\begin{algorithm}
\caption{ Search for a Starting Point $\bm{\theta}_0$}
\label{alg_sivartheta}
\begin{algorithmic}[1]
\STATE \textbf{Initialization:}  $l\leftarrow 1$;
\REPEAT
\STATE let $(\hat{q},\hat{p})$ be the index of the largest entry in $\mathbf{y}$ by amplitude;
\STATE $\phi_l \leftarrow \bar{\phi}_{\hat{p}}$, $\psi_l\leftarrow \bar{\psi}_{\hat{q}}$, $\mathbf{h}_l\leftarrow \big(\mathbf{F}^T\otimes \mathbf{W}^H\big) \big(\mathbf{e}^*_t(\cos{\bar{\phi}_{\hat{p}}})\otimes\mathbf{e}_r(\cos{\bar{\psi}_{\hat{q}}})\big)$, $\alpha_l\leftarrow \mathbf{h}_l^H\mathbf{y}/(\mathbf{h}_l^H\mathbf{h}_l)$;
\STATE $\mathbf{y}\leftarrow \mathbf{y}-\mathbf{h}_l\alpha_l$;
\STATE $l\leftarrow l+1$;
\UNTIL{$\alpha_l<\alpha_{\text{th}}$}
\STATE{remove the last entries in $\bm{\phi}$, $\bm{\psi}$, and $\bm{\alpha}$, respectively;}
\STATE $\bm{\theta}_0 \leftarrow [\bm{\phi}^T \bm{\psi}^T]^T$.
\end{algorithmic}
\end{algorithm}

With Algorithm \ref{alg_sivartheta}, we obtain an initial estimate of $\bm{\alpha}$ and $\bm{\theta}$, then we use the vector $\bm{\theta}_0$ as the starting point for solving problem (\ref{eqn_lsrealvartheta}) with the LM algorithm. It is worth mentioning that Algorithm \ref{alg_sivartheta} is essentially the same with the adaptive estimation method for the single RF chain case in \cite{alkhateeb2014cehpmwcs} and similar to the OMP algorithm in \cite{lee2014essechmsmwc}. Unlike those methods which require a large number of quantization levels to achieve  good performance, our estimation accuracy relies mainly on whether the obtained starting point is within the range of a good local optimal point. With a good starting point, the LM algorithm would improve the estimation accuracy. Hence, our method requires much fewer pilots than their methods to achieve the same performance.

\subsection{Calculation of the Jacobian matrix}

The Jacobian matrix $\mathbf{J}$ in (\ref{eqn_lswc}) consists of $2L$ columns:
\begin{align}
\mathbf{J}=[\frac{\partial{\mathbf{r}(\bm{\theta})}}{\partial{\phi_1}}, \ldots, \frac{\partial{\mathbf{r}(\bm{\theta})}}{\partial{\phi_L}}, \frac{\partial{\mathbf{r}(\bm{\theta})}}{\partial{\psi_1}}, \ldots, \frac{\partial{\mathbf{r}(\bm{\theta})}}{\partial{\psi_L}} ].
\end{align}
By (\ref{eqn_rfunc}),
\begin{align} \label{eqn_lthjacob}
\frac{\partial{\mathbf{r}(\bm{\theta})}}{\partial{\phi_l}}=-\frac{\partial(\bm{\Phi}(\bm{\theta})\bm{\Phi}^\dag(\bm{\theta}))}{\partial\phi_l}\mathbf{y}
=-\Big(\frac{\partial\bm{\Phi}(\bm{\theta})}{\partial\phi_l}\bm{\Phi}^\dag(\bm{\theta})+\bm{\Phi}(\bm{\theta})\frac{\partial\bm{\Phi}^\dag(\bm{\theta})}{\partial\phi_l}\Big)\mathbf{y}.
\end{align}
Here $\frac{\partial\bm{\Phi}(\bm{\theta})}{\partial\phi_l}$ is the matrix obtained by taking the derivative of each element in $\bm{\Phi}(\bm{\theta})$ w.r.t. $\phi_l$. $\frac{\partial\bm{\Phi}^\dag(\bm{\theta})}{\partial\phi_l}$ is also obtained by taking the derivative of each element of $\bm{\Phi}^\dag(\bm{\theta})$ w.r.t. $\phi_l$, and can be using $\frac{\partial\bm{\Phi}(\bm{\theta})}{\partial\phi_l}$ \cite{pseudoinverse}:
\begin{align} \label{eqn_partialpseudo}
\frac{\partial\bm{\Phi}^\dag(\bm{\theta})}{\partial\phi_l}=&-\bm{\Phi}^\dag(\bm{\theta})\frac{\partial\bm{\Phi}(\bm{\theta})}{\partial\phi_l}\bm{\Phi}^\dag(\bm{\theta})
+\bm{\Phi}^\dag(\bm{\theta})\bm{\Phi}^{\dag T}(\bm{\theta})\frac{\partial\bm{\Phi}^T(\bm{\theta})}{\partial\phi_l}\big(I-\bm{\Phi}(\bm{\theta})\bm{\Phi}^\dag(\bm{\theta})\big)\\
&+\big(I-\bm{\Phi}^\dag(\bm{\theta})\bm{\Phi}(\bm{\theta})\big)\frac{\partial\bm{\Phi}^T(\bm{\theta})}{\partial\phi_l}\bm{\Phi}^{\dag T}(\bm{\theta})\bm{\Phi}^\dag(\bm{\theta}).\nonumber
\end{align}

Since only the elements in the $l$th column of $\bm{\Phi}(\bm{\theta})$ depend on variable $\phi_l$, all but the $l$th column in $\frac{\partial\bm{\Phi}(\bm{\theta})}{\partial\phi_l}$ are zero vectors. The $(q+(p-1)m_r)$th element in the $l$th column of $\bm{\Phi}(\bm{\theta})$ is $G^{qp}_l(\bm{\theta})$, and its derivative  w.r.t. $\phi_l$ is given as 
\begin{align} \label{eqn_partialG1}
\frac{\partial{G^{qp}_l(\bm{\theta})}}{\partial{\phi_{l}}}=&\frac{-\sin\phi_{l}}{n_r n_t}\frac{1-e^{-j\pi n_r\Omega_r^{lq}}}{1-e^{-j\pi\Omega_r^{lq}}}\frac{j\pi e^{j\pi \Omega_t^{lp}}-j\pi n_t e^{j\pi n_t \Omega_t^{lp}}+j\pi(n_t-1)e^{j\pi(n_t+1)\Omega_t^{lp}}}{(1-e^{j\pi\Omega_t^{lp}})^2},
\end{align}
where $\Omega_t^{lp}=\cos{\phi_l}-\cos{\bar{\phi}_p}$ and $\Omega_r^{lq}=\cos{\psi_l}-\cos{\bar{\psi}_q}$.

The expression of $\frac{\partial{\mathbf{r}(\bm{\theta})}}{\partial{\psi_l}}$ is similar as (\ref{eqn_lthjacob}), with $\phi_l$ substituted by $\psi_l$. Again, only the $l$th column in $\frac{\partial\bm{\Phi}(\bm{\theta})}{\partial\psi_l}$ is nonzero, and the $(q+(p-1)m_r)$th element in this column is given as 
\begin{align} \label{eqn_partialG2}
\frac{\partial{G^{qp}_l(\bm{\theta})}}{\partial{\psi_{l}}}=&\frac{-\sin\psi_{l}}{n_r n_t}\frac{-j\pi e^{-j\pi \Omega_r^{lq}}+j\pi n_r e^{-j\pi n_r \Omega_r^{lq}}-j\pi(n_r-1) e^{-j\pi(n_r+1)\Omega_r^{lq}}}{(1-e^{-j\pi\Omega_r^{lq}})^2}\frac{1-e^{j\pi n_t\Omega_t^{lp}}}{1-e^{j\pi\Omega_t^{lp}}}.
\end{align}

The preceding seemingly complicated formulas are in fact straightforward to compute. With the starting point obtained by Algorithm \ref{alg_sivartheta} and the algorithm for calculating the Jacobian matrix, we can apply the LM algorithm for estimating the channel parameters. However, since $\mathbf{r}(\bm{\theta})$ and $\mathbf{J}$ in (\ref{eqn_lswc}) are complex, by directly applying LM to (\ref{eqn_lswc}) would result in complex values for $\bm{\theta}$. To deal with this, let
$\tilde{\mathbf{r}}(\bm{\theta})=\left[\begin{array}{c}
\text{Re}(\mathbf{r}(\bm{\theta}))\\
\text{Im}(\mathbf{r}(\bm{\theta}))
\end{array}\right ],
$
$
\tilde{\mathbf{J}}=\left[\begin{array}{c}
\text{Re}(\mathbf{J})\\
\text{Im}(\mathbf{J})
\end{array}\right],
$
then (\ref{eqn_lswc}) can be equivalently formulated as
\begin{align} \label{eqn_lswc_ref}
\tilde{\mathbf{r}}(\bm{\theta})\approx\tilde{\mathbf{r}}(\bm{\theta}_{k-1})+ \tilde{\mathbf{J}}(\bm{\theta}_{k-1})(\bm{\theta}-\bm{\theta}_{k-1}), \quad ||\mathbf{S}\bm{\theta}||_2\leq \delta.
\end{align}
By applying LM algorithm on (\ref{eqn_lswc_ref}), $\bm{\theta}$ is always real in each iteration. To summarize, we have Algorithm \ref{alg_LMCE} for channel acquisition.

\begin{algorithm}
\caption{LM based Channel Acquisition}
\label{alg_LMCE}
\begin{algorithmic}[1]
\STATE \textbf{Initialization:} $\mathbf{y}$;
\STATE Use Alogrithm \ref{alg_sivartheta} to obtain an initial estimate of $\bm{\alpha}_0$, $\bm{\theta}_0$;
\STATE Use $\bm{\theta}_0$ as the starting point for the LM algorithm;
\STATE Use the LM algorithm for Problem (\ref{eqn_lswc_ref}), and get the optimal point $\bm{\theta}$;
\STATE $\bm{\alpha}\leftarrow \bm{\Phi}^\dag(\bm{\theta})\mathbf{y} $;
\STATE Output $\bm{\alpha}$ and $\bm{\theta}$.
\end{algorithmic}
\end{algorithm}

Algorithm \ref{alg_LMCE} is a heuristic method for solving problem (\ref{eqn_lsrealvartheta}), and there is no guarantee that the obtained solution is global optimal. However, it exhibits a fairly good estimation performance in the simulations. Besides, Algorithm \ref{alg_LMCE} provides a general way for channel estimation, and it does not rely on the channel variation model. Although it can be employed to track the CSI, it is more suitable for CSI acquisition.

\section{Channel tracking} \label{sec_chtrack}

In Subsection \ref{subsec_chavarmod}, we assume that the path gain vector $\bm{\alpha}$ does not change while the channel angle vector $\bm{\theta}$ evolves by  (\ref{eqn_anglevary}) within a block. With the acquisition of $\bm{\alpha}$ and $\bm{\theta}$ using Algorithm \ref{alg_LMCE} at the beginning of each block, it is only necessary to keep tracking the angle vector $\bm{\theta}$ within the block. The channel variation model (\ref{eqn_anglevary}) and the observation (\ref{eqn_vecobserve}) fit into the Kalman filter framework except for the nonlinearity of $\bm{\Phi}(\bm{\theta})$ w.r.t. $\bm{\theta}$ ($\bm{\alpha}$ is known here after acquisition since we assume it does not change within a block). Though the signal evolution model (\ref{eqn_anglevary}) is linear, the observation vector $\mathbf{y}$ is nonlinear w.r.t. the angle vector $\bm{\theta}$ in (\ref{eqn_vecobserve}), we should first use a linear approximation for the observation, given as follows
\begin{align} \label{eqn_linearappro}
\mathbf{y}(n)&=\bm{\Phi}(\bm{\theta}(n))\bm{\alpha}+\mathbf{v}(n) \\
&=\bm{\Phi}(\bm{\theta}(n)-\bm{\hat{\theta}}(n|n-1)+\bm{\hat{\theta}}(n|n-1))\bm{\alpha}+\mathbf{v}(n)\\
&\approx \bm{\Phi}(\bm{\hat{\theta}}(n|n-1))\bm{\alpha}+  \mathbf{C}(n)\Big(\bm{\theta}(n)-\bm{\hat{\theta}}(n|n-1)\Big)+\mathbf{v}(n)\\
&=\mathbf{C}(n)\bm{\theta}(n)+\mathbf{v}(n)+\mathbf{d}(n)
\end{align}
where $\bm{\hat{\theta}}(n|n-1)$ is the linear minimum mean square error prediction of $\bm{\theta}(n)$ based on $\{\mathbf{y}(0), \mathbf{y}(1),\ldots, \mathbf{y}(n-1)\}$, the $l$th column ($1\leq l\leq L$) in $\mathbf{C}(n)$ is $\frac{\partial\bm{\Phi}(\bm{\theta}(n))}{\partial \phi_l(n)}|_{\phi_l(n)=\hat{\phi}_l(n|n-1)}\bm{\alpha}$ and for columns from $L+1$ to $2L$, $\phi_l(n)$ is substituted by $\psi_l(n)$, and
\begin{align}
\mathbf{d}(n)=\bm{\Phi}(\bm{\hat{\theta}}(n|n-1))\bm{\alpha}-\mathbf{C}(n)\bm{\hat{\theta}}(n|n-1).
\end{align}

Directly applying Kalman filter on (\ref{eqn_anglevary}) and (\ref{eqn_linearappro}) would result in complex values for the angles $\bm{\theta}$, we transform complex vectors and matrices into real-valued ones. Let
\begin{align}
\tilde{\mathbf{y}}(n)=\left[\begin{array}{c}
\text{Re}(\mathbf{y}(n))\\
\text{Im}(\mathbf{y}(n))
\end{array} \right], &
\tilde{\mathbf{C}}(n)=\left[\begin{array}{c}
\text{Re}(\mathbf{C}(n)) \\
\text{Im}(\mathbf{C}(n))
\end{array}\right],
\tilde{\mathbf{v}}(n)=\left[\begin{array}{c}
\text{Re}(\mathbf{v}(n)) \\
\text{Im}(\mathbf{v}(n))
\end{array}\right], &
\tilde{\mathbf{d}}(n)=\left[\begin{array}{c}
\text{Re}(\mathbf{d}(n))\\
\text{Im}(\mathbf{d}(n))
\end{array} \right].
\end{align}
Then (\ref{eqn_linearappro}) can be equivalently formulated as
\begin{align} \label{eqn_linearappro_real}
\tilde{\mathbf{y}}(n)\approx\tilde{\mathbf{C}}(n)\bm{\theta}(n)+\tilde{\mathbf{v}}(n)+\tilde{\mathbf{d}}(n).
\end{align}

Based on (\ref{eqn_linearappro_real}), the algorithm is shown as Algorithm \ref{alg_kfct}. The parameters used in the algorithm are:  $\widehat{\bm{\theta}}(n|i)$ denotes the linear mean square error estimator of $\bm{\theta}(n)$ based on $\{\tilde{\mathbf{y}}(0), \tilde{\mathbf{y}}(1),\ldots, \tilde{\mathbf{y}}(i)\}$, $\mathbf{M}(n|n-1)=\mathbb{E}\big((\bm{\theta}(n)-\widehat{\bm{\theta}}(n|n-1))(\bm{\theta}(n)-\widehat{\bm{\theta}}(n|n-1))^H\big)$  denotes the minimum prediction mean square error (MSE) matrix, $\mathbf{M}(n|n)=\mathbb{E}\big((\bm{\theta}(n)-\widehat{\bm{\theta}}(n|n))(\bm{\theta}(n)-\widehat{\bm{\theta}}(n|n))^H\big)$ denotes the minimum MSE matrix, $\mathbf{K}(n)$ is the Kalman gain matrix, and $\widehat{\bm{\theta}}(0|0)$, $\mathbf{M}(0|0)$ are the initial values of $\widehat{\bm{\theta}}(n|n), \mathbf{M}(n|n)$, respectively.

\begin{algorithm}
\caption{Kalman Filter based Channel Tracking}
\label{alg_kfct}
\begin{algorithmic}[1]
\STATE \textbf{Initialization:} $\widehat{\bm{\theta}}(0|0)\leftarrow\bm{\theta}(0), \mathbf{M}(0|0)\leftarrow \mathbf{0}$, $\bm{\alpha}(0)$;
\WHILE{slot $n$ still in the same block}
\STATE Prediction:
 $\widehat{\bm{\theta}}(n|n-1) \leftarrow \widehat{\bm{\theta}}(n-1|n-1)$;
\STATE Minimum Prediction MSE Matrix:
 $\mathbf{M}(n|n-1) \leftarrow \mathbf{M}(n-1|n-1)+\mathbf{Q}_u$;
\STATE Kalman Gain Matrix:
$\mathbf{K}(n) \leftarrow \mathbf{M}(n|n-1)\tilde{\mathbf{C}}^H(n)\Big(\mathbf{Q}_v+\tilde{\mathbf{C}}(n)\mathbf{M}(n|n-1)\tilde{\mathbf{C}}^H(n)\Big)^{-1}$;
\STATE Correction:
$\widehat{\bm{\theta}}(n|n)\leftarrow \widehat{\bm{\theta}}(n|n-1)+\mathbf{K}(n)\Big(\tilde{\mathbf{y}}(n)-\bm{\Phi}\big(\widehat{\bm{\theta}}(n|n-1)\big)\bm{\alpha}(n)\Big)$;
\STATE \textbf{Output $\widehat{\bm{\theta}}(n|n)$ as the estimate of $\bm{\theta}(n)$};
\STATE Minimum MSE Matrix:
$\mathbf{M}(n|n)\leftarrow \Big(\mathbf{I}-\mathbf{K}(n)\tilde{\mathbf{C}}(n)\Big)\mathbf{M}(n|n-1)$;
\STATE $n\leftarrow n+1$;
\ENDWHILE
\end{algorithmic}
\end{algorithm}

Algorithm \ref{alg_kfct} provides a recursive way to estimate the channel parameters, so each time we only need to store the current observation vector $\tilde{\mathbf{y}}(n)$ and the previous estimation $\widehat{\bm{\theta}}(n-1|n-1)$.

In practice, we do not know the variation noise covariance matrix  $\mathbf{Q}_u$, but as proven in \cite{reif1999ssdtekf}, so long as $\mathbf{Q}_u\succeq \xi \mathbf{I}$, where $\succeq$ is the generalized inequality defined on the positive semidefinite cone, $\xi \in \mathcal{R}_+$, and other criteria are satisfied, the estimation error of the extended Kalman filter is exponentially bounded and bounded with probability $1$. So knowing the exact value of $\mathbf{Q}_u$ is not necessary in Algorithm \ref{alg_kfct}, and we can simply let $\mathbf{Q}_u= \xi \mathbf{I}$.

\section{Abrupt Change Detection} \label{sec_acd}

In this section, we develop an algorithm to detect abrupt channel changes. At any time $n$, the two hypotheses are:

$\mathcal{H}_0$: there is no abrupt change in the channel;

$\mathcal{H}_1$: there are abrupt changes in the channel.

We use $\mathbf{y}(n)$ for such a test and define
\begin{align} \label{eqn_abtchdet}
L(\mathbf{y}(n))&= \big(\mathbf{y}(n)-\bm{\Phi}(\hat{\bm{\theta}}(n))\hat{\bm{\alpha}}(n)\big)^H \mathbf{Q}_v^{-1} \big(\mathbf{y}(n)-\bm{\Phi}(\hat{\bm{\theta}}(n))\hat{\bm{\alpha}}(n)\big),
\end{align}
where $\hat{\bm{\alpha}}(n)$ and $\hat{\bm{\theta}}(n)$ are the estimate of $\bm{\alpha}$ and $\bm{\theta}$ using Kalman filter (Algorithm \ref{alg_kfct}) respectively. If the channel acquisition and tracking are accurate enough, $L(\mathbf{y}(n))$ would be small in the case without abrupt changes since the term remained in $\mathbf{y}(n)-\bm{\Phi}(\hat{\bm{\theta}}(n))\hat{\bm{\alpha}}(n)$ is noise according to (\ref{eqn_vecobserve}). On the other hand, upon any abrupt changes, the difference includes gains of some dominant path. We then decide $\mathcal{H}_1$ if
\begin{align} \label{eqn_Lgamma}
L(\mathbf{y}(n))>\gamma,
\end{align}
where $\gamma$ is a predefined threshold and should be chosen to be high enough to meet the desired false alarm probability, defined as
\begin{align}\label{eqn_fa}
P_{\text{FA}}=P\{L(\mathbf{y}(n))>\gamma;\mathcal{H}_0\}.
\end{align}
An approximate expression for $\gamma$ given a $P_{\text{FA}}$ can be developed as follows If the acquisition of channel parameters in the beginning of each block and the tracking in each slot are all accurate enough, $2L(\mathbf{y}(n))$ would follow Chi-Squared distribution with degree $2m_tm_r$ (since it is the sum of the square of $2m_tm_r$ normal random variables), i.e., $2L(\mathbf{y}(n))\sim \chi_{2m_tm_r}^2$, where $\chi_{\nu}^2$ denotes Chi-Squared distribution with degree $\nu$. Let $Q_{\chi_{\nu}^2}(x)$ be the right-tail probability for a $\chi_{\nu}^2$ random variable, then given a false alarm probability $P_{\text{FA}}$, we can determine a corresponding threshold
\begin{align} \label{eqn_thres}
\gamma=\frac{1}{2}Q_{\chi_{2m_rm_t}^2}^{-1}(P_{\text{FA}}).
\end{align}
We will show that this approximate threshold works well for the proposed framework in the simulations.

The detection method is presented as Algorithm \ref{alg_GLRTacd}.
\begin{algorithm}
\caption{Abrupt Change Detection}
\label{alg_GLRTacd}
\begin{algorithmic}[1]
\STATE Given a $P_{\text{FA}}$, determine the threshold $\gamma$ using (\ref{eqn_thres});
\WHILE{True}
\STATE In slot $n$, use Algorithm \ref{alg_kfct} to obtain $\hat{\bm{\theta}}(n)$;
\STATE Determine whether there are abrupt changes based on (\ref{eqn_abtchdet}) and (\ref{eqn_Lgamma});
\ENDWHILE
\end{algorithmic}
\end{algorithm}

\section{Beamforming and combing vector selection} \label{sec_sysscheme}

The overall proposed system is shown in Fig. \ref{fig_gen_procedure}, where Algorithms \ref{alg_LMCE}-\ref{alg_GLRTacd} are devised for channel acquisition, tracking, and abrupt change detection, respectively.

Here, each pilot corresponds to a pair of pilot beamforming and combing vectors $(\bar{\phi}_p,\bar{\psi}_q)$. In the proposed system, either for the channel acquisition method or the channel tracking method, we use the pilots to sample the range of AoDs and AoAs. In order to ensure detection of abrupt changes of paths with any AoD and AoA, it is best that the number of pilots should be no smaller than $n_tn_r$, or equivalently, the minimum number of beamforming and combining directions are equal to the number of antennas at the transmitter and receiver, respectively. The reason is that to detect changes of all paths, for any $(\phi,\psi)\in [0,\pi]\times[0,\pi]$, there should be at least one pilot $(\bar{\phi}_p,\bar{\psi}_q)$ with large enough beamforming and combing gain $|\mathbf{e}_r^H(\bar{\psi}_{q})\mathbf{e}_r(\psi)\mathbf{e}_t^H(\phi)\mathbf{e}_t(\bar{\phi}_{p})|$. Otherwise, we cannot detect changes of paths with weak beamforming and combining gains for all pilots using Algorithm \ref{alg_GLRTacd}.

\begin{figure}
  \centering
  \includegraphics[width=0.8\columnwidth]{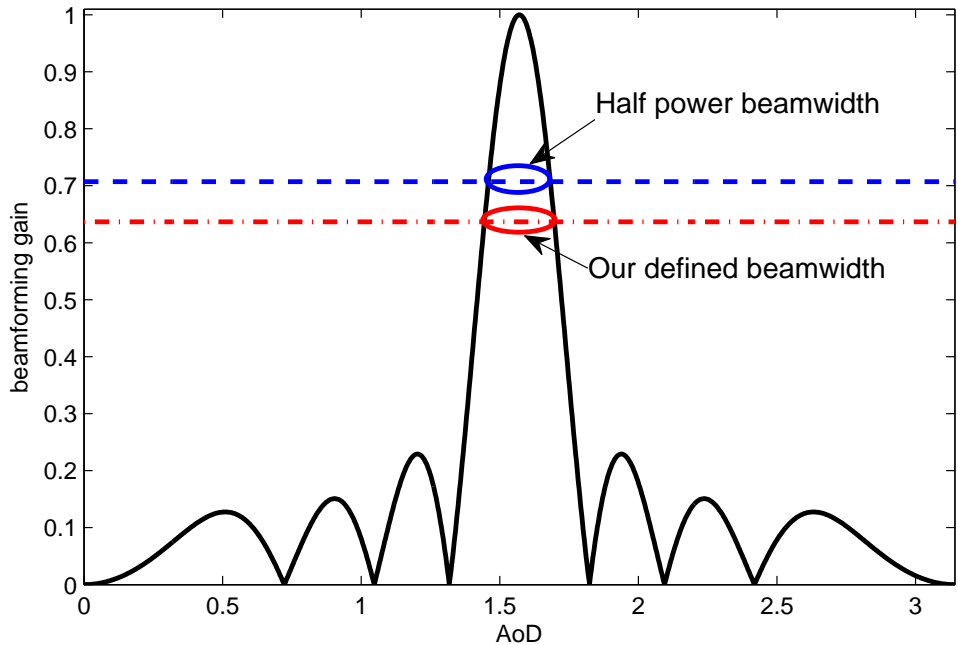}\\
  \caption{Main lobe and sidelobes for $\bar{\phi}_p=\frac{\pi}{2}$, $n_t=8$.}\label{fig_mainsideLobe}
\end{figure}

To see how many pilots are necessary to guarantee this, we take a closer look at the beamforming gain $|\mathbf{e}_t^H(\phi)\mathbf{e}_t(\bar{\phi}_{p})|$, and the combining gain is similar. From  (\ref{eqn_fsample}), we obtain
\begin{align} \label{eqn_beamgain}
|\mathbf{e}_t^H(\phi)\mathbf{e}_t(\bar{\phi}_{p})|=\bigg|\frac{\sin\big(\frac{\pi}{2}n_t(\cos{\phi}-\cos{\bar{\phi}_p})\big)}{n_t\sin\big(\frac{\pi}{2}(\cos{\phi}-\cos{\bar{\phi}_p})\big)}\bigg|.
\end{align}
For a fixed $\bar{\phi}_p$, we see from (\ref{eqn_beamgain}) that there is a main lobe around $\bar{\phi}_p$ such that $|\cos{\phi}-\cos{\bar{\phi}}_p|\leq \frac{2}{n_t}$, and any other lobes are weak compared to this main lobe, as shown by Fig. \ref{fig_mainsideLobe}. In antenna theory, the half power beamwidth is often used as the beamwidth of the main lobe, i.e., the angles between the points of the main lobe where power is half of its maximum. We can also define the beamwidth as such, however, for ease of computation, we define the beamwidth of $\bar{\phi}_p$ as the width between the two points that $|\cos{\phi}-\cos{\bar{\phi}}_p|=\frac{1}{n_t}$ \cite[Chap. 7]{tse2005fowc}. It can be calculated that when $n_t$ is large, the beamforming gain at these two points is approximately $|\mathbf{e}_t^H(\phi)\mathbf{e}_t(\bar{\phi}_{p})|=\frac{2}{\pi}$, and the power reduces to approximately $0.4053$ of the maximum power, which is still a reasonable power for detection.

We use each beamforming direction $\bar{\phi}_p$ to cover the angles within its beamwidth, then we can determine the required number of beamforming directions (number of quantization levels). However, since the angles are in cosine form in the expression of the beamforming gain (\ref{eqn_beamgain}), the beamwidths of different $\bar{\phi}_p$ are different, so we can not obtain the number by dividing $\pi$ by the beamwidth of one $\bar{\phi}_p$. Instead, since $\cos{\phi}\in[-1,1]$, and the beamwidths of all $\bar{\phi}_p$ in the cosine form are $\frac{2}{n_t}$, the minimum number of beamforming directions is $m_t^{\text{min}}=n_t$.

\begin{figure}
  \centering
  \includegraphics[width=0.8\columnwidth]{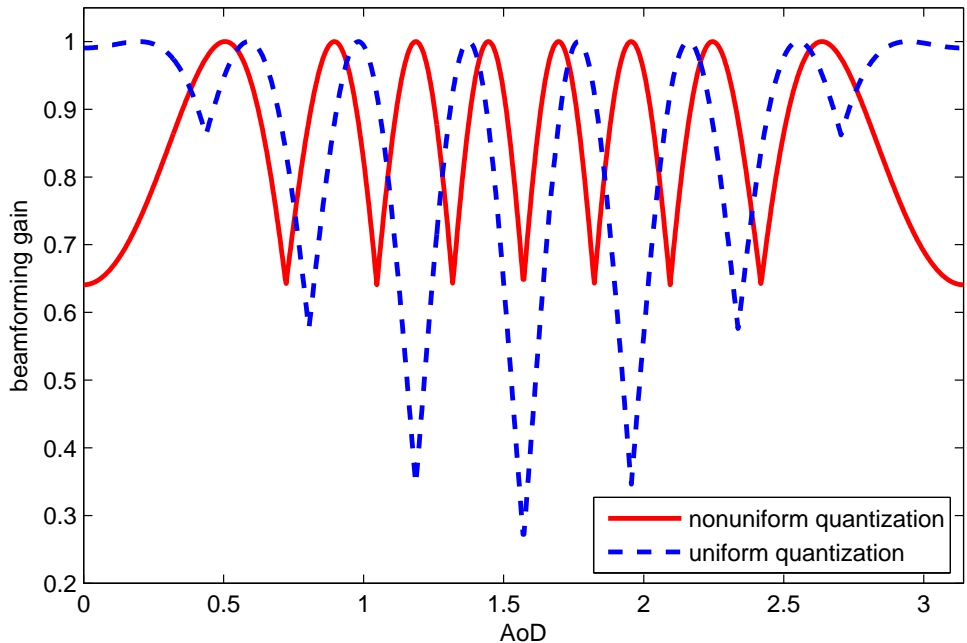}\\
  \caption{Beamforming directions for $m_t=n_t=8$. }\label{fig_beamdirections}
\end{figure}

To determine the beamforming directions, we can use the following method. Suppose $m_t\geq n_t$ beamforming directions are available, we divide the range of $\cos{\phi}$, i.e., $[-1,1]$ into $m_t$ bins, and then use the arccosine of the center of each bin as the beamforming direction. This quantization is nonuniform for the angles. The beamforming directions are plotted for the case where $m_t=n_t=8$ in Fig. \ref{fig_beamdirections}, where each beamforming direction corresponds to the point with maximum beamforming gain $1$. For comparison, we also plot the results of uniform quantization commonly employed by previous works, which divides $[0,\pi]$ uniformly and then uses the center of each bin as the beamforming direction. As can be seen from this figure, for the same number of quantization levels, uniform quantization has areas with pretty weak beamforming gain, for instance, around $1.6$, the gain can be smaller than $0.3$. On the other hand, the nonuniform quantization method ensures all angles have beamforming gains no smaller than $0.63$.

For channel acquisition, the number of quantization levels at both transmitters and receivers should be no smaller than $n_t$ and $n_r$, respectively, to ensure all paths can be estimated. For channel tracking, the number can be much smaller than $n_t$ or $n_r$. However, it is difficult to obtain a threshold for the tracking method. In the simulations, we find that even with quantization levels $\frac{n_t}{2}$ and $\frac{n_r}{2}$ for the transmitter and receiver, respectively, the tracking performance remains quite good. For detecting abrupt changes, again, the minimum requirement should be satisfied. Considering that we need to determine whether there are changes in each slot, we use $n_t$, $n_r$ quantization levels for beamforming and combining, respectively, even for channel tracking, and use numbers slightly larger than or equal to $n_t$ and $n_r$, respectively, for channel acquisition.

\section{Simulation Results} \label{sec_simulation}

\begin{table}
\centering
\caption{Common Simulation parameters}
\label{tab_compars}
\begin{tabular}{c|c|c|c|c|c|c|c}
  \Xhline{1.2pt}
  \Gape[1.5ex]{$n_t$} & $n_r$ & slot length & $\mathbf{Q}_u$ & $\mathbf{Q}_v$  \\
  \hline
   \Gape[1.5ex]{16} & 16 & 0.1 ms & $\sigma_u^2\mathbf{I}_{2L}$ & $\sigma_v^2\mathbf{I}_{m_rm_t}$  \\
  \Xhline{1.2pt}
\end{tabular}
\end{table}

To evaluate either the acquisition or tracking performance, we define the normalized mean square error (NMSE) of $\mathbf{H}$,
\begin{align} \label{eqn_NMSE}
\epsilon(\mathbf{H},\hat{\mathbf{H}})=10\log_{10}\Big(\frac{\mathbb{E}[||\hat{\mathbf{H}}-\mathbf{H}||_F^2]}{\mathbb{E}[||\mathbf{H}||_F^2]}\Big),
\end{align}
as the metric, where $\hat{\mathbf{H}}$ is the estimated channel matrix, $\mathbf{H}$ is the true channel matrix, and $||*||_F$ takes the Frobenius norm of a matrix. Conceivably, the smaller $\epsilon(\mathbf{H},\hat{\mathbf{H}})$ is, the better the estimation performance. Besides, SNR used in the following is defined as $\text{SNR}=10\log_{10}\big(\frac{n_tn_r}{\sigma_v^2}\big)$, where we normalize the transmit power as $1$. Common simulation parameters for all cases are given in Table \ref{tab_compars}.

\begin{figure}
  \centering
  \includegraphics[width=0.8\columnwidth]{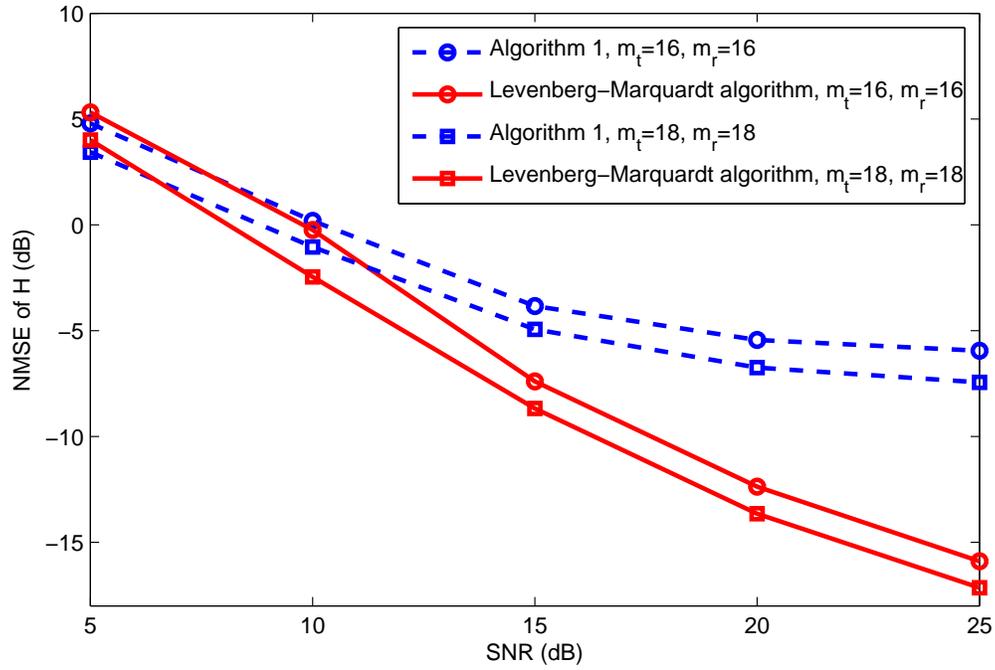}\\
  \caption{ NMSE of $\mathbf{H}$ versus SNR for channel acquisition. }\label{fig_chacqnmsevssnr}
\end{figure}

\begin{figure}
  \centering
  \includegraphics[width=0.8\columnwidth]{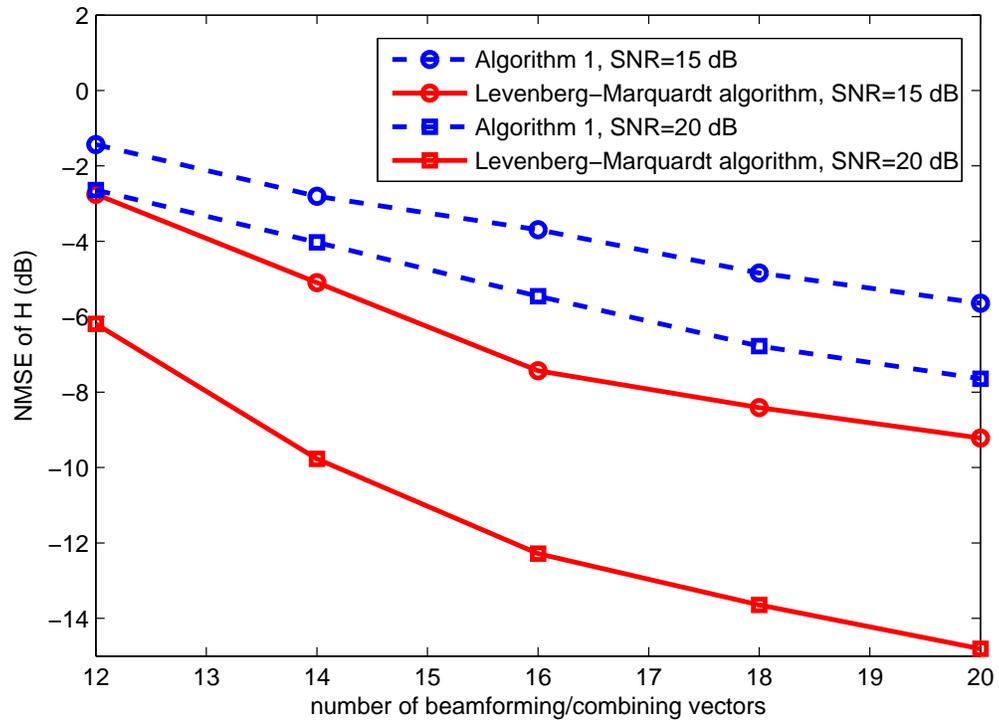}\\
  \caption{ NMSE of $\mathbf{H}$ versus number of quantization levels for channel acquisition. }\label{fig_chacqnmsevsquant}
\end{figure}

\subsection{Channel acquisition performance}

We first show the channel acquisition performance of the LM algorithm (Algorithm \ref{alg_LMCE}). For comparison, we use Algorithm \ref{alg_sivartheta}, which has similar performance with algorithms proposed in \cite{alkhateeb2014cehpmwcs} and \cite{lee2014essechmsmwc}.  Fig. \ref{fig_chacqnmsevssnr} and Fig. \ref{fig_chacqnmsevsquant} show how both algorithms are influenced by the SNR and the number of quantization levels (the required number of pilots is the multiplication of numbers of quantization levels at the transmitter and the receiver), respectively. In both figures, we simulate $1000$ slots for each point, and in each slot, we generate $L=3$ paths independently, with the path gains following $\mathcal{CN}(0,n_tn_r)$ and the AoDs and AoAs following uniform distribution on $(0, \pi)$. For the search algorithm, we estimate $5$ paths, while for the LM algorithm we first estimate $5$ paths and then only keep paths with SNR larger than $10$ dB. So the actual number of estimated paths with this algorithm can be smaller than $5$. Fig. \ref{fig_chacqnmsevssnr} shows that the estimation accuracy of both algorithms would improve with the increase of SNR. It also indicates that when the SNR is low, i.e., less than $10$ dB, both algorithms have very poor acquisition performance. The reason is that the least squares solution minimizes the residual with noise in the observation (\ref{eqn_lsori}), when the power of noise is large, the solution may be far from the one minimizing the residual without noise in the observation (which essentially is $||\hat{\mathbf{H}}-\mathbf{H}||_F^2$). When the SNR is relatively high, the LM algorithm shows significant improvement over the search algorithm.

Fig. \ref{fig_chacqnmsevsquant} indicates that the increase of beamforming/combing vectors could also improve performances of both algorithms. As discussed in \ref{sec_sysscheme}, in general, the number of beamforming/combing vectors should be no smaller than the number of antennas at the transmitter and receiver, respectively; otherwise, some paths might lie in the range that none of the pilots has large enough beamforming/combining gain to detect them. If the number of quantization levels is low, the advantage of the LM algorithm over the search algorithm would diminish.

\begin{figure}
  \centering
  \includegraphics[width=0.8\columnwidth]{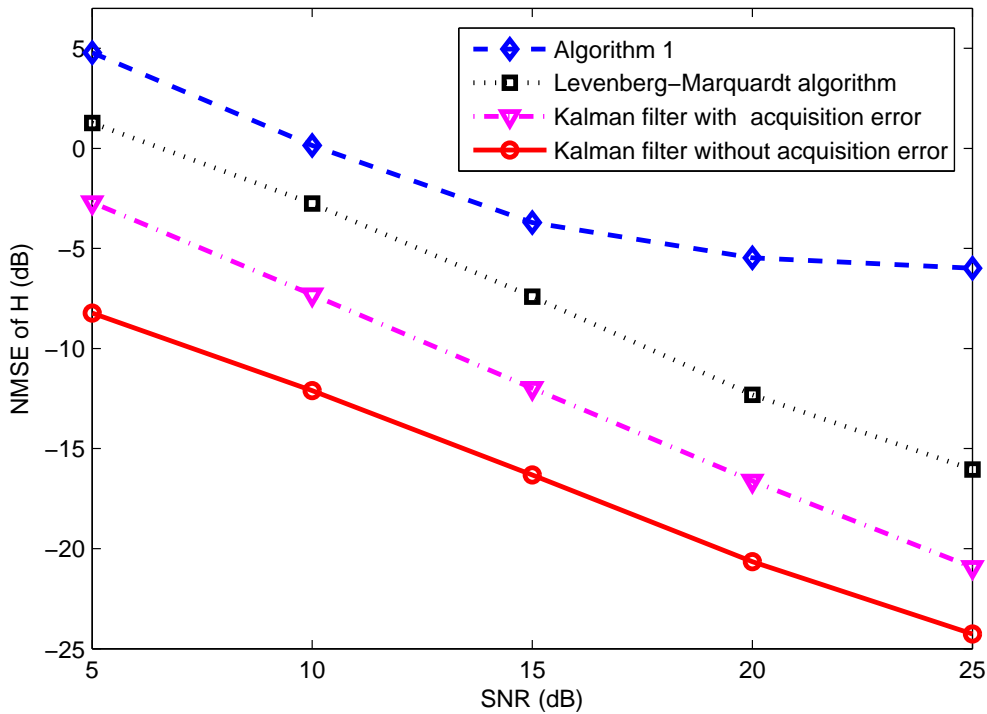}\\
  \caption{ NMSE of $\mathbf{H}$ versus SNR for channel tracking, $m_t=16$, $m_r=16$, $\sigma_u^2=(\frac{0.5}{180}\pi)^2$. }\label{fig_chtrnmsevssnr}
\end{figure}

\begin{figure}
  \centering
  \includegraphics[width=0.8\columnwidth]{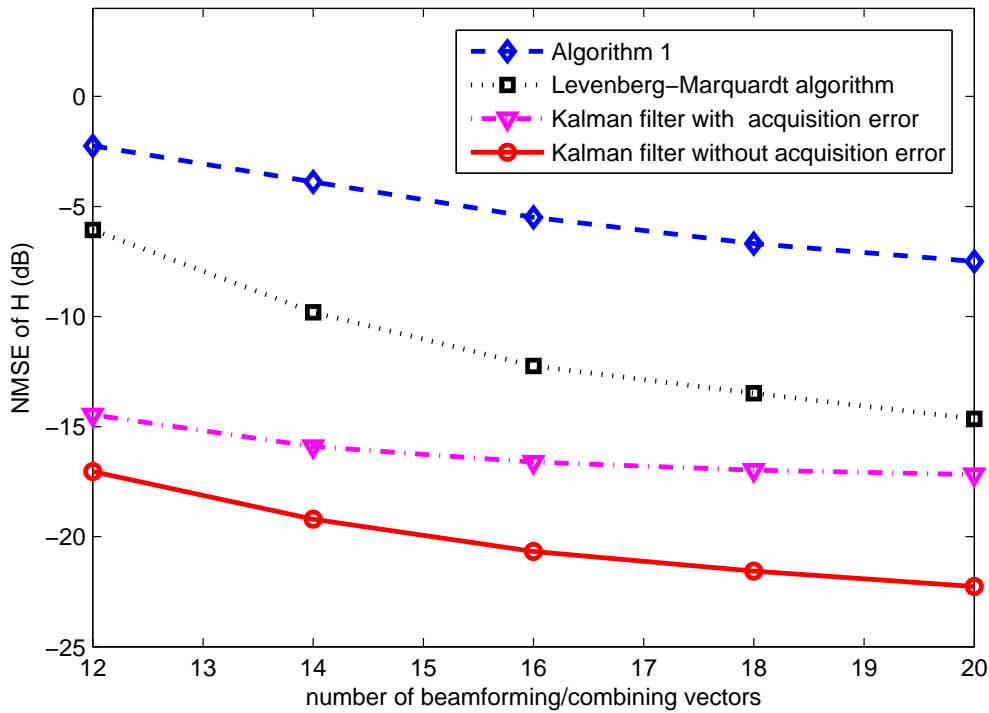}\\
  \caption{ NMSE of $\mathbf{H}$ versus number of beamforming/combining vectors for channel tracking, SNR=$20$ dB, $\sigma_u^2=(\frac{0.5}{180}\pi)^2$. }\label{fig_chtrnmsevsquant}
\end{figure}

\begin{figure}
  \centering
  \includegraphics[width=0.8\columnwidth]{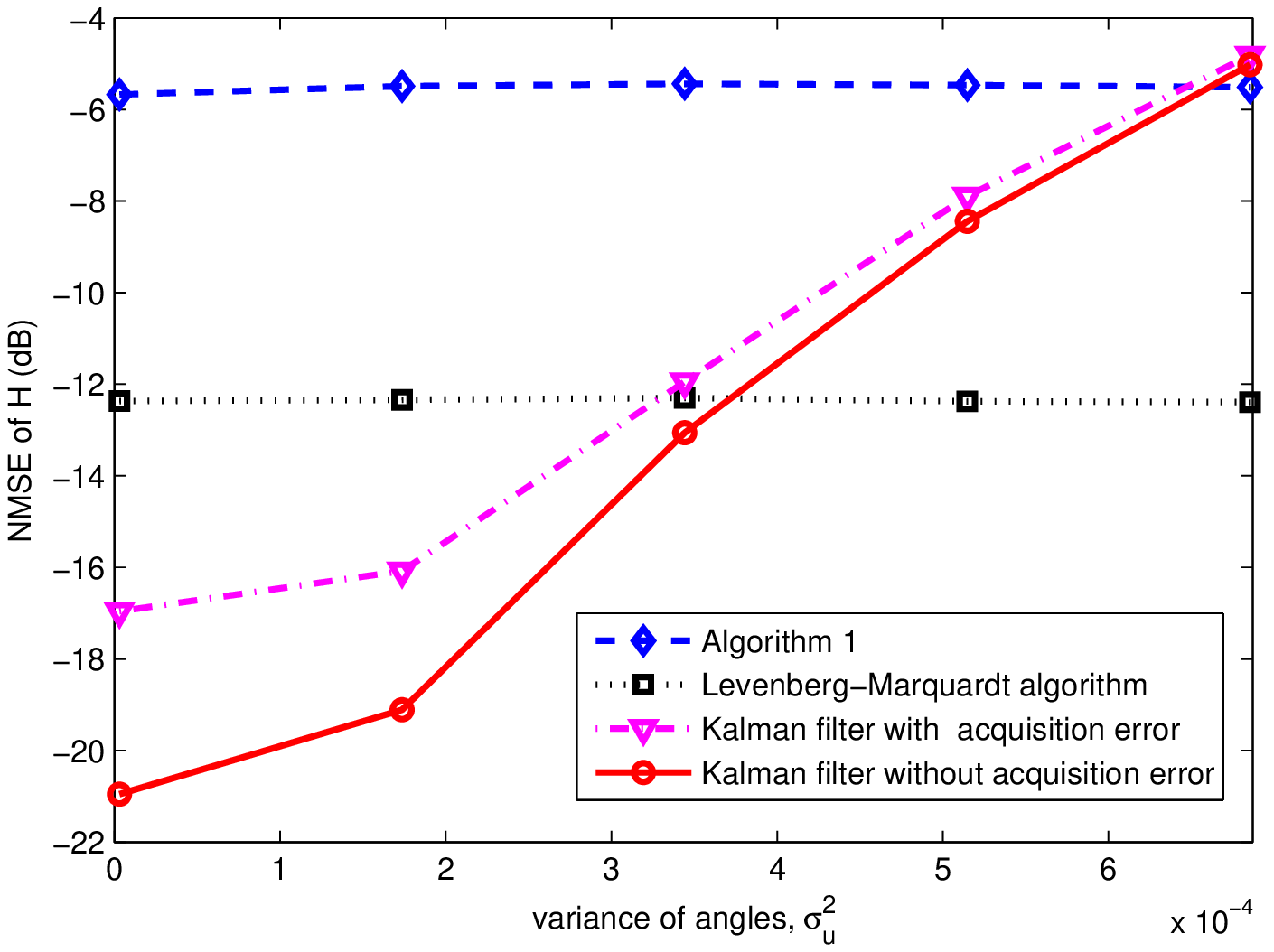}\\
  \caption{ NMSE of $\mathbf{H}$ versus channel variation speed for channel tracking, $\text{SNR}=20$ dB, $m_t=16$, $m_r=16$.  }\label{fig_chtrnmsevsvar}
\end{figure}

\subsection{Channel tracking performance}

We then compare the performance of the Kalman filter algorithm (Algorithm \ref{alg_kfct}) with that of the LM algorithm (Algorithm \ref{alg_LMCE}) and the search algorithm (Algorithm \ref{alg_sivartheta}) for channel tracking. For ease of simulation, we set $\mathbf{Q}_u=\sigma_u^2\mathbf{I}_{2L}$ in Table \ref{tab_compars}, which means that variations of AoDs, AoAs of different paths are independent, the proposed algorithm applies to the general correlated case as well. We can change $\sigma_u^2$ to simulate different channel variation speeds. Since the initial values of $\bm{\alpha}$ and $\bm{\theta}$ also have an impact on the tracking performance, we simulate $1000$ blocks with different initial values. In each block, there are $50$ slots, and in the first slot, we generate $L=3$ paths with their gains following $\mathcal{CN}(0,n_tn_r)$ and angles according to uniform distribution on $(0,\pi)$. Then in the following slots, the path gains do not change while the angles evolve according to (\ref{eqn_anglevary}). For the Kalman filter algorithm, we simulate two cases, one without acquisition error, and the other with  acquisition error (a randomly generated noise vector following $\mathcal{CN}(0,\sigma^2_v\mathbf{I}_L)$ is added to the path gain vector $\bm{\alpha}$). The other two algorithms estimate the path gain vector and the angle vector in each slot. The channel variation speed parameter $\sigma_u^2$ is unknown to all three algorithms. In the Kalman filter algorithms, we use $(\frac{2}{180}\pi)^2$ as the assumed value of $\sigma_u^2$ instead of its real value. The other two algorithms do not need this parameter.

In Figs. \ref{fig_chtrnmsevssnr}-\ref{fig_chtrnmsevsvar}, we compare their tracking performance in terms of SNR, number of quantization levels and channel variation speed, respectively. As shown by Figs. \ref{fig_chtrnmsevssnr} and \ref{fig_chtrnmsevsquant}, when the variation speed is $\sigma_u^2=(\frac{0.5}{180}\pi)^2$, the Kalman filter algorithm has higher tracking accuracy than the other two, even with acquisition error. This also implies the possibility of jointly considering channel acquisition and tracking.

However, there are two disadvantages with the Kalman filter algorithm. One is shown in Fig. \ref{fig_chtrnmsevsvar}, its tracking performance would deteriorate with the increase of the channel variation speed, which mainly arises from applying Kalman filter for nonlinear observation. When the variation noise is large, like $3.5\times 10^{-4}$ in the figure, the Kalman filter algorithm loses track, and estimation error can accumulate gradually. In the simulations, we find that when $\sigma_u^2\leq(\frac{1}{180}\pi)^2$, the Kalman filter algorithm has pretty accurate tracking performance. We can roughly analyze how fast the angle varies with this noise variance. The average drift of the angle ($\mathbb{E}(|u|)$) would be $\sqrt{\frac{2}{\pi}}\sigma_u\approx 0.0139$ rad, suppose each slot lasts $0.1$ ms, then the angle variation speed is $7964^\circ/s$, which is equivalent to rotating a mobile device more than $22$ rounds per second. The other disadvantage of this algorithm is that it requires an initial accurate estimate of channel parameters. If the estimate is not accurate enough, its performance can get much worse. In the following simulations, we would utilize the LM algorithm for the initial estimate.

\begin{figure*}
  \centering
  \includegraphics[width=\columnwidth]{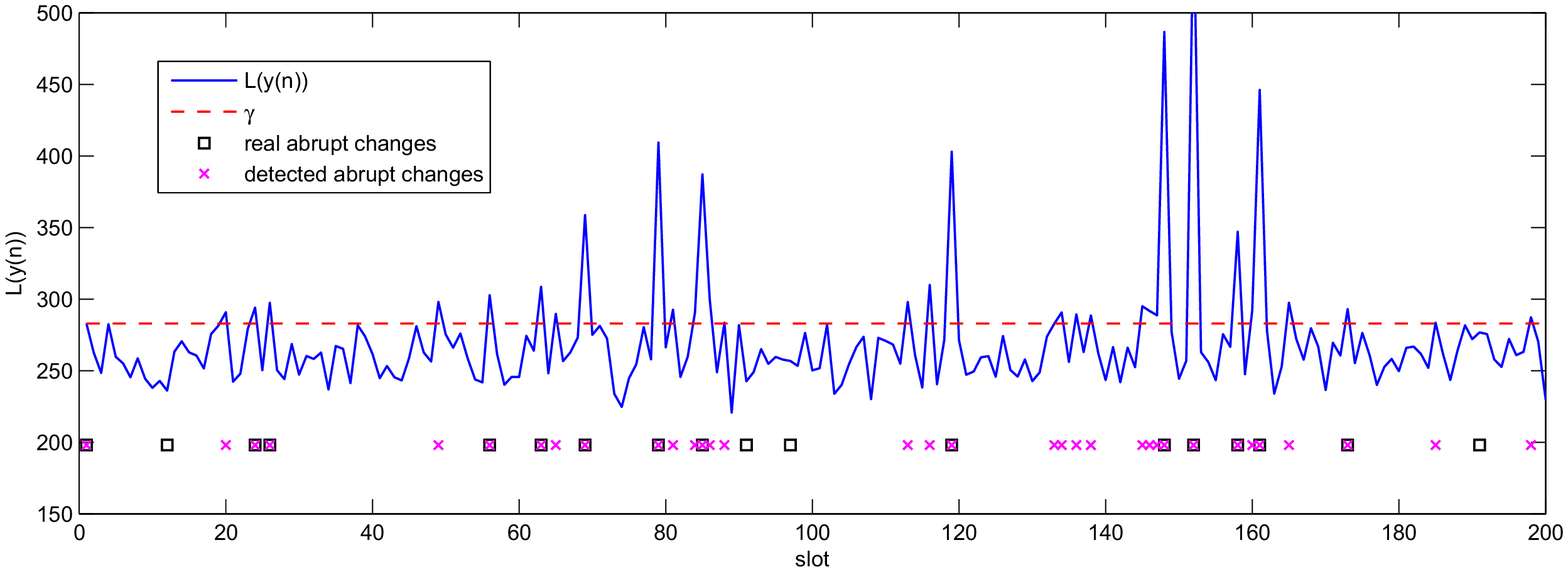}\\
  \caption{ Abrupt change detection, $m_t=16$, $m_r=16$, SNR=$20$ dB, $p_{\text{FA}}=0.05$, $\sigma_u^2=(\frac{0.5}{180}\pi)^2$. }\label{fig_GLRT}
\end{figure*}

\begin{figure*}
  \centering
  \includegraphics[width=\columnwidth]{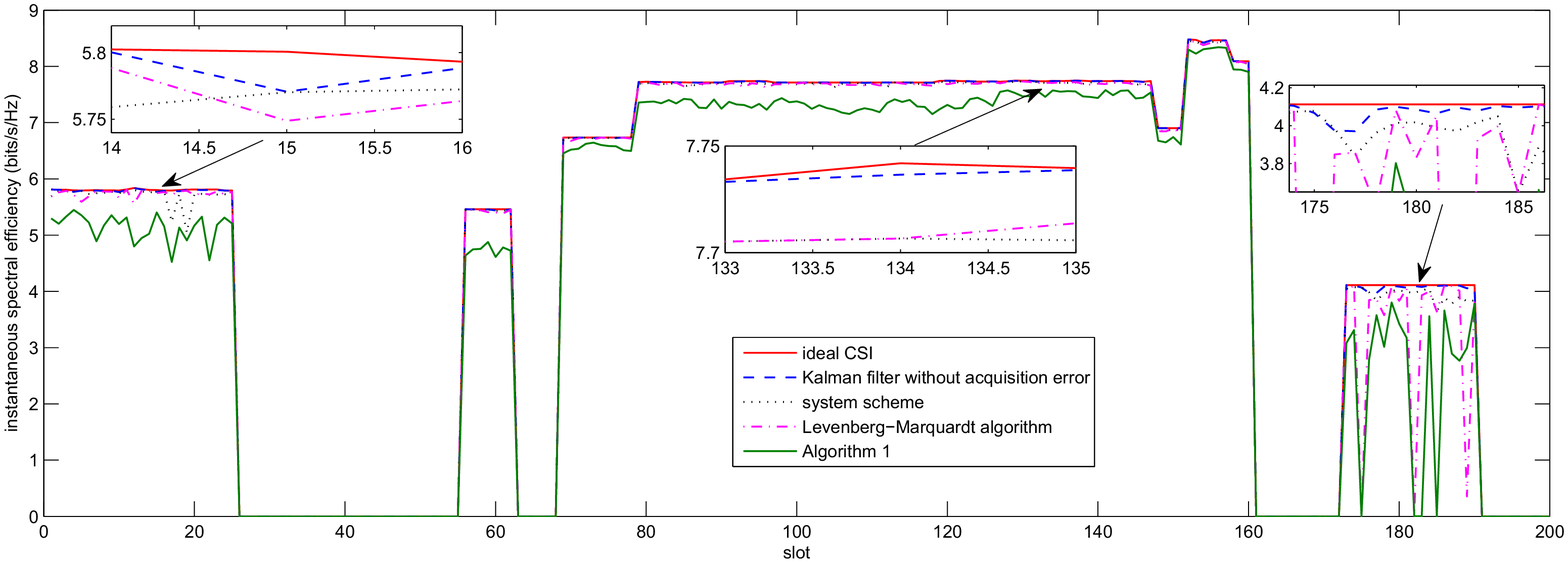}\\
  \caption{ Instantaneous spectral efficiency, $m_t=16$, $m_r=16$, SNR=$20$ dB, $p_{\text{FA}}=0.05$, $\sigma_u^2=(\frac{0.5}{180}\pi)^2$. }\label{fig_spec_effi}
\end{figure*}

\subsection{Performance of integrated scheme}

Finally, we compare the performances of the combined channel acquisition, tracking and abrupt change detection scheme (denoted as system scheme) with that of using only the search algorithm, the LM algorithm and the Kalman filter algorithm without acquisition error for a mm-wave channel with both abrupt changes and slow variations. We simulate the channel in the following way. In the first slot, we generate $L=3$ paths randomly as before. In the following slots, the arrivals and departures of paths form an  M/M/$\infty$ queue, where the arrival rate is $\lambda=500$ times/second and the departure rate is $\mu=200$ times/second. Suppose each slot lasts $0.1$ ms, then the average duration of a path is $50$ slots. The disappearance of a path is done by setting the path gain to be zero, while the appearance is by generating its gain following $\mathcal{CN}(0, n_tn_r)$ and angles uniformly on $(0, \pi)$. Within a block, the path gain vector of existing paths does not change while the angle vector evolves according to (\ref{eqn_anglevary}).

To show how well each scheme works as well as to reflect how channel varies, we use the spectral efficiency achieved by each scheme instead of the NMSE of $\mathbf{H}$ as the metric. The spectral efficiency is calculated in the following way. Based on the estimated channel matrix $\hat{\mathbf{H}}$, we calculate the optimal beamforming and combining vector by solving the following optimization problem,
\begin{align} \label{eqn_fw_est}
\max_{||\mathbf{f}||=1,||\mathbf{w}||=1}& \quad |\mathbf{w}^H\hat{\mathbf{H}}\mathbf{f}|,
\end{align}
which is essentially the induced 2 norm of $\hat{\mathbf{H}}$. Then by using the optimal $\hat{\mathbf{f}}$ and $\hat{\mathbf{w}}$ of (\ref{eqn_fw_est}) in the real channel, the spectral efficiency is
\begin{align} \label{eqn_spec_effi}
R=\log_2\Big(1+\frac{|\hat{\mathbf{w}}^H\mathbf{H}\hat{\mathbf{f}}|^2}{\sigma_v^2}\Big).
\end{align}

It is worth mentioning that $\hat{\mathbf{f}}$ and $\hat{\mathbf{w}}$ cannot be realized by the single RF chain transceiver in our scheme since the elements may not be of constant modulus. However, we still use them because if we restrict $\hat{\mathbf{f}}$ and $\hat{\mathbf{w}}$ to be of constant modulus, the spectral efficiency of all schemes are very close. Since we focus on the channel estimation performance, we relax the beamforming/combining schemes in order to reveal more details of the estimated channel. Besides, when the devised scheme is extended to transceivers with multiple RF chains, the same results would apply for them as well.

Fig. \ref{fig_GLRT} shows the abrupt change detection results. The blue solid curve is the evolution of the log likelihood ratio $L(\mathbf{y}(n))$ and the red dashed line is the threshold set for the test. In addition, we use black squares to denote the slots when real abrupt changes occur and magenta x-mark to denote the slot when an abrupt change is declared by the detector. As can be seen, most abrupt changes are detected except for the slots $12$, $91$, $97$ and $191$, which are due to paths with minor gains. Besides, there are $19$ slots that false alarm occur in $200$ slots. This is almost twice as high as the predefined false alarm probability. The reason is due to the acquisition error and the accumulation of tracking error.

Fig. \ref{fig_spec_effi} plots the instantaneous spectral efficiency versus time curves of ideal CSI, Kalman filter without acquisition error, system scheme, LM algorithm and search algorithm. It reveals that both the system scheme and the LM algorithm can achieve spectral efficiency very close to the case with ideal CSI, and the reduction is in most cases within $0.1$ bits/s/Hz in the enlarged figures. The search algorithm always has the lowest spectral efficiency due to inaccurate channel estimation, which can be $1$ bits/s/Hz lower than the ideal CSI case in some slots.

To illustrate the overhead of the scheme more clearly, we assume that each slot lasts $0.1$ ms, and the symbol rate is $20$ million symbols per second. Based on the above discussions, with $n_t=n_r=16$, the numbers of pilots required by channel acquisition, tracking and change detection are all $256$, so in each slot, there is only $12.8\%$ overhead of pilots. In addition, we need to feedback the CSI regularly in each slot. The key difference between channel acquisition and tracking is the amount of feedback information, if there is no abrupt change, we only need to feedback the angle vector $\bm{\theta}$ obtained by channel tracking, while if abrupt changes are detected, we need to feedback both the channel gain vector $\bm{\alpha}$ and the angle vector $\bm{\theta}$ obtained by channel acquisition.

\section{Conclusion} \label{sec_conclusion}

In this paper, we have proposed a framework for estimating mm-wave channels with both abrupt changes and slow variations. By treating abrupt changes and slow variations differently, we are able to reduce the overhead of pilots and computation by decomposing the channel estimation problem into three parts, i.e., channel acquisition, tracking, and abrupt change detection. We have developed an algorithm for each part and discussed the design of pilots and implementations of the proposed scheme. Simulation results indicate that the proposed scheme estimates the mm-wave channel accurately under moderate SNR. Compared with existing estimation methods, the proposed scheme takes channel variations into account, and it answers the question of how often pilots should be sent implicitly. Moreover, the proposed scheme is based on analog beamforming/combining and requires only a few number of quantization levels for phase shifters, thus it is quite suitable for practical implementation.

\bibliographystyle{IEEEtran}
\bibliography{mmwave}

\begin{thebibliography}{10}
\providecommand{\url}[1]{#1}
\csname url@samestyle\endcsname
\providecommand{\newblock}{\relax}
\providecommand{\bibinfo}[2]{#2}
\providecommand{\BIBentrySTDinterwordspacing}{\spaceskip=0pt\relax}
\providecommand{\BIBentryALTinterwordstretchfactor}{4}
\providecommand{\BIBentryALTinterwordspacing}{\spaceskip=\fontdimen2\font plus
\BIBentryALTinterwordstretchfactor\fontdimen3\font minus
  \fontdimen4\font\relax}
\providecommand{\BIBforeignlanguage}[2]{{%
\expandafter\ifx\csname l@#1\endcsname\relax
\typeout{** WARNING: IEEEtran.bst: No hyphenation pattern has been}%
\typeout{** loaded for the language `#1'. Using the pattern for}%
\typeout{** the default language instead.}%
\else
\language=\csname l@#1\endcsname
\fi
#2}}
\providecommand{\BIBdecl}{\relax}
\BIBdecl

\bibitem{chuang2016tadammwc}
C.~Zhang, D.~Guo, and P.~Fan, ``Tracking angles of departure and arrival in a
  mobile millimeter wave channel,'' in \emph{Proc. IEEE Int. Conf. Commun.},
  May 2016.

\bibitem{boccardi2014fdtd5g}
F.~Boccardi, R.~W. {Heath Jr.}, A.~Lozano, T.~L. Marzetta, and P.~Popovski,
  ``Five disruptive technology directions for {5G},'' \emph{IEEE Commun. Mag.},
  vol.~52, no.~2, pp. 74--80, Feb. 2014.

\bibitem{andrews2014ww5gb}
J.~G. Andrews, S.~Buzzi, W.~Choi, S.~V. Hanly, A.~Lozano, A.~C.~K. Soong, and
  J.~C. Zhang, ``What will 5g be?'' \emph{IEEE J. Sel. Areas Commun.}, vol.~32,
  no.~6, pp. 1065 -- 1082, Jun. 2014.

\bibitem{rappaport2013mwmc5gc}
T.~Rappaport, S.~Sun, R.~Mayzus, H.~Zhao, Y.~Azar, K.~Wang, G.~Wong, J.~Schulz,
  M.~Samimi, and F.~Gutierrez, ``Millimeter wave mobile communications for 5g
  cellular: It will work!'' \emph{IEEE Access}, vol.~1, pp. 335--349, Sep.
  2013.

\bibitem{pi2011aimwmbs}
Z.~Pi and F.~Khan, ``An introduction to millimeter-wave mobile broadband
  systems,'' \emph{IEEE Commmun. Mag.}, vol.~49, no.~6, pp. 101--107, Jun.
  2011.

\bibitem{rappaport2015wideband}
T.~S. Rappaport, G.~R. MacCartney, M.~K. Samimi, and S.~Sun, ``Wideband
  millimeter-wave propagation measurements and channel models for future
  wireless communication system design,'' \emph{IEEE Transactions on
  Communications}, vol.~63, no.~9, pp. 3029--3056, Sept 2015.

\bibitem{samimi201628}
M.~K. Samimi, G.~R. MacCartney, S.~Sun, and T.~S. Rappaport, ``28 {GHz}
  millimeter-wave ultrawideband small-scale fading models in wireless
  channels,'' in \emph{2016 IEEE 83rd Vehicular Technology Conference (VTC
  Spring)}, May 2016, pp. 1--6.

\bibitem{samimi2016mimo}
M.~K. Samimi, S.~Sun, and T.~S. Rappaport, ``{MIMO} channel modeling and
  capacity analysis for {5G} millimeter-wave wireless systems,'' in \emph{2016
  10th European Conference on Antennas and Propagation (EuCAP)}, April 2016,
  pp. 1--5.

\bibitem{degli2014ray-tracing-based}
V.~Degli-Esposti, F.~Fuschini, E.~M. Vitucci, M.~Barbiroli, M.~Zoli, L.~Tian,
  X.~Yin, D.~A. Dupleich, R.~M¨¹ller, C.~Schneider, and R.~S.
  Thom$\ddot{\text{a}}$, ``Ray-tracing-based mm-wave beamforming assessment,''
  \emph{IEEE Access}, vol.~2, pp. 1314--1325, 2014.

\bibitem{ayach2014sspmwms}
O.~E. Ayach, S.~Rajagopal, S.~Abu-Surra, Z.~Pi, and R.~W. {Heath Jr.},
  ``Spatially sparse precoding in millimeter wave {MIMO} systems,'' \emph{IEEE
  Trans. Wireless Commun.}, vol.~13, no.~3, pp. 1499--1513, Mar. 2014.

\bibitem{akdeniz2014mmwcmcce}
M.~R. Akdeniz, Y.~Liu, M.~K. Samimi, S.~Sun, S.~Rangan, T.~S. Rappaport, and
  E.~Erkip, ``Millimeter wave channel modeling and cellular capacity
  evaluation,'' \emph{IEEE J. Sel. Areas Commun.}, vol.~32, no.~6, pp. 1164 --
  1179, Jun. 2014.

\bibitem{alkhateeb2014cehpmwcs}
A.~Alkhateeb, O.~E. Ayach, G.~Leus, and R.~W. {Heath Jr.}, ``Channel estimation
  and hybrid precoding for millimeter wave cellular systems,'' \emph{IEEE J.
  Sel. Topics Signal Process.}, vol.~8, no.~5, pp. 831--846, Oct. 2014.

\bibitem{lee2014essechmsmwc}
J.~Lee, G.~T. Gil, and Y.~H. Lee, ``Exploiting spatial sparsity for estimating
  channels of hybrid {MIMO} systems in millimeter wave communications,'' in
  \emph{Proc. IEEE Global Commun. Conf.}, Dec. 2014, pp. 3326--3331.

\bibitem{tse2005fowc}
D.~Tse and P.~Viswanath, \emph{Fundamentals of Wireless Communication}.\hskip
  1em plus 0.5em minus 0.4em\relax Cambridge University Press, 2005.

\bibitem{hur2013mmbwbascn}
S.~Hur, T.~Kim, D.~J. Love, J.~V. Krogmeier, T.~A. Thomas, and A.~Ghosh,
  ``Millimeter wave beamforming for wireless backhaul and access in small cell
  networks,'' \emph{IEEE Trans. Commun.}, vol.~61, no.~10, pp. 4391 -- 4403,
  Oct. 2013.

\bibitem{kay2009estimation}
S.~M. Kay, \emph{Fundamentals of Statistical Signal Processing, Volume I:
  Estimation Theory}.\hskip 1em plus 0.5em minus 0.4em\relax Prentice Hall PTR,
  Jan. 2009.

\bibitem{levenberg1944amscnpls}
K.~Levenberg, ``A method for the solution of certain nonlinear problems in
  least squares,'' \emph{Quart. Appl. Math.}, vol.~2, pp. 164 -- 168, 1944.

\bibitem{mor1978LMait}
J.~J. Mor$\acute{\text{e}}$, ``The {Levenberg-Marquardt} algorithm:
  Implementation and theory,'' \emph{Numerical Analysis, G.A. Watson, ed.,
  Springer-Verlag}, pp. 105 -- 116, 1978.

\bibitem{laub2004mafse}
A.~J. Laub, \emph{Matrix Analysis for Scientists and Engineers}.\hskip 1em plus
  0.5em minus 0.4em\relax Philadelphia, PA, USA: Society for Industrial and
  Applied mathematics, 2004.

\bibitem{pseudoinverse}
``{Moore-Penrose pseudoinverse},''
  https://en.wikipedia.org/wiki/Moore

\bibitem{reif1999ssdtekf}
K.~Reif, S.~G$\ddot{\text{u}}$nther, E.~Yaz, and R.~Unbehauen, ``Stochastic
  stability of the discrete-time extended {Kalman} filter,'' \emph{IEEE Trans.
  Automat. Contr.}, vol.~44, no.~4, pp. 714 -- 728, Apr. 1999.

\end{thebibliography}

\end{document}